\documentclass{ws06-procs}

\begin{document}

\title{Measures Of Solar Oscillations And Supergranulation By The Magnetic-Optical
Filter}

\author{Ilaria Formicola\footnote{e-mail: ilariaformicola@libero.it}, Andrea Longobardo\footnote{e-mail: andrea.longobardo@tele2.it}, Ciro Pinto\footnote{e-mail: ciropinto1982@libero.it},\\ Pierluigi Cerulo\footnote{e-mail: lizard031@hotmail.it}}

\address{University of Naples Federico II, Complesso Universitario di Monte Sant'Angelo, Via Cinthia, I-80126, Naples, Italy}

\maketitle

\abstracts{Helioseismology is the branch of the Solar Physics
which studies the solar global oscillations, a phenomenon very
important in order to understand the inside of the Sun. We explain
here a method for measuring the solar velocity fields along the
line of sight by the VAMOS instrument developed at the
Astronomical Observatory of Capodimonte in Naples, in particular
we present the measures of the so called five-minutes oscillations
and of the Supergranulation}

Helioseismology is the branch of the Solar Physics which studies
the solar global oscillations. From the power spectrum we can
obtain informations about the solar inside, for example the
convective zone's depth, the inner layers' rotation velocities,
the temperature, the density and the chemical composition. The
success of Helioseismology has carried this branch to be extended
to analogous studies in other stars (Asteroseismology). An
important role in this branch is played by the VAMOS (Velocity And
Magnetic Observations of the Sun) project, developed in Naples, at
the Astronomical Observatory of Capodimonte. The employed
instrument allows us to scan solar photosphere's images in
intensity, velocity field and  magnetic field along the line of
sight; in our work, we used the second type of images.

The VAMOS instrument is composed essentially of a Magnetic-Optical
Filter (MOF) and a Wing Selector (WS). MOF consists of a potassium
vapours cell with a magnetic field (about 1400 G) along its optic
axis, interposed between two crossed linear polarizers. In order
to understand how this works, we have to  recall the Zeeman
effect. Let's consider the atomic transition from the level with
$l = 1$ to that with $l = 0$ (where $l$ is the angular momentum
quantum number): in absence of magnetic field, there is only an
emission line. If we are in presence of a magnetic field, the
degeneration of the level with $l = 1$ is removed bringing to
three different states with three different values of the atomic
quantum number $m$ (magnetic moment quantum number) and we can see
no more {\itshape{one}} emission line, but {\itshape {three}}
emission lines characterized by different states of polarization.
In fact two of these emission lines are circularly polarized,
respectively right-handed ($\sigma^+$) and left-handed
($\sigma^-$), around the magnetic field direction, the other
($\pi$) is linearly polarized along the magnetic field, so, when
we observe along this direction (it's our case) we can't see this
last component. MOF is based on two effects: the Righi Effect and
the Macaluso-Corbino Effect. Righi Effect is Zeeman effect in
absorption: solar light (not polarized) arrives on the first
polarizer which transforms it in linearly polarized light (let's
recall that linearly polarized light can be viewed as half right
circularly polarized and half left circularly polarized); then the
cell absorbs half of  the light intensity at $\sigma^+$ and
$\sigma^-$ wavelengths and the second polarizer cuts half of the
light intensity at $\sigma^+$ and $\sigma^-$ wavelengths and cuts
totally the other wavelengths. So, at the output, we should see
only two peaks at the Zeeman wavelengths, but the net output of
the filter is characterized by the presence of the
Macaluso-Corbino Effect, too. This consists in a rotation of the
polarization plane caused by a difference in refraction index
values at the two Zeeman wavelengths in the cell. Higher is the
temperature of the cell, stronger is the Macaluso-Corbino Effect
which shows itself as two additional symmetric peaks, the distance
between which increases linearly with temperature (in the range
considered in our work).

WS is positioned after MOF and its role is to select only one of
the two MOF output lines. It is composed by a quarter-wave plate
and a cell analogous to MOF's cell. If the plate's transmission
axis forms with the optic axis an angle of 45 degrees, light's
polarization becomes right circular and so the  cell cuts the
$\sigma^+$ component, while leaves the $\sigma^-$ one to pass; if
the plate's transmission axis forms with the optic axis an angle
of -45 degrees, we have the opposite situation and only the
$\sigma^-$ component passes.

A velocity image of the Sun is called {\itshape {dopplergram}}.
There are many contributes in a dopplergram: some vary very slowly
during an observation time (15-20 minutes) and so they can be
considered constant, other, instead, vary faster. Contributes that
we considered constant are caused by Earth's proper motions
(revolution and rotation), solar rotation and gravitational
redshift. Instead, contributes quickly variable are due to solar
oscillations, granulation, supergranulation and noise. Except the
noise, we wanted to measure these. In reality, we have obtained
estimations of solar oscillations and supergranulation's velocity,
but not of granulation's one because VAMOS spatial resolution
doesn't allow to observe this phenomenon.

Solar oscillations (or {\itshape {five minutes}} oscillations)
were observed for the first time by Leighton in 1962, but the
first theoretical models, in agreement with observations, were
developed in the 70's. These oscillations are mainly vertical (so
more visible towards the center of the solar image), have an
amplitude of 500 m/s, periods of about five minutes and lifetimes
that vary from few hours to months. Five minutes oscillations are
due to wave propagating under the photosphere. There are
essentially two types of waves: acoustic waves (p-modes),
generated by pressure variations caused by convective
instabilities, and gravity waves (g-modes), transversal, caused by
gas' stratification and generated where there are density
discontinuities. Gravity waves aren't directly observable, except
surface waves (f-modes), propagating in the low photosphere. To
measure five minutes oscillations, we considered two dopplergrams,
obtained at a temporal interval of two minutes and half (half
period). Semidifference between the two dopplergrams gives us the
maximum amplitude of solar oscillations. We repeated this
proceeding for other two images and then we obtained an average,
the {\itshape{difference-dopplergram}}, which gives us, with a
good estimation, the five minutes oscillations' profile.

Supergranulation is the photospherical evidence of the Solar
convection. Supergrains are 35 times bigger than grains (and so we
can observe them with the VAMOS spatial resolution), their mean
lifetime is one day  and velocities, in this case prevalently
horizontal, go from 300 to 500 m/s. To measure supergranulation,
we considered once again two dopplergrams obtained at a temporal
interval of two minutes and half, but in this case we did a
semisum rather than a semidifference: in fact, in this way, we
excluded the solar oscillations' contributes. The remaining
contributes are due to supergranulation and constant motions. We
obtained constant motions spatially smoothing the {\itshape
{sum-dopplergram}} image; indeed, by subtracting the smoothed
dopplergram to the former, we could obtain the dopplergram
relative to the supergranulation.

All the operations on dopplergrams have been made using the IDL
program language.

\end{document}